# "Wigner crystal" and "stripe" models for the magnetic and crystallographic superstructures of $La_{0.333}Ca_{0.667}MnO_3$.

P.G. Radaelli

*ISIS Facility, Rutherford Appleton Laboratory, Bldg. R3, Chilton, Didcot Oxon. OX11 0QX UK*

D.E. Cox

*Department of Physics, Brookhaven National Laboratory, Upton, NY 11973*

L. Capogna

*Institut Laue Langevin, B.P. 156 F. 38042 Grenoble France and INFM, Via Perrone, Genova Italy*

S-W. Cheong

*Bell Laboratories, Lucent Technologies, Murray Hill, NJ 07974 and Department of Physics and Astronomy, Rutgers University Piscataway, New Jersey, 08855*

M. Marezio

*MASPEC-CNR, via Chiavari 18A, 43100 Parma, Italy*

# "Wigner crystal" and "stripe" models for the magnetic and crystallographic superstructures of $La_{0.333}Ca_{0.667}MnO_3$.

P.G. Radaelli
*ISIS Facility, Rutherford Appleton Laboratory, Bldg. R3, Chilton, Didcot Oxon. OX11 0QX UK*

D.E. Cox
*Department of Physics, Brookhaven National Laboratory, Upton, NY 11973*

L. Capogna
*Institut Laue Langevin, B.P. 156 F. 38042 Grenoble France and INFM, Via Perrone, Genova Italy*

S-W. Cheong
*Bell Laboratories, Lucent Technologies, Murray Hill, NJ 07974 and Department of Physics and Astronomy, Rutgers University Piscataway, New Jersey, 08855*

M. Marezio
*MASPEC-CNR, via Chiavari 18A, 43100 Parma, Italy*

ABSTRACT

The crystallographic (charge-ordered) and magnetic superstructures of $La_{0.333}Ca_{0.667}MnO_3$ were studied by high-resolution synchrotron x-ray and neutron powder diffraction. In the antiferromagnetic structure, which was refined using a non-collinear model, the a lattice parameter is *tripled* and the c lattice parameter is *doubled* with respect to the average crystallographic unit cell (Pnma setting). The crystallographic structure below the charge-ordering temperature ($T_{CO} \sim 260$ K) is characterized by ordering of the $d_{z^2}$ orbitals of the Jahn-Teller-distorted $Mn^{3+}O_6$ octahedra in the orthorhombic ac plane, and the appearance of superlattice peaks in the x-ray patterns corresponding to a *tripling* of the a axis lattice parameter. The intensities of the superlattice peaks can be accounted for satisfactorily in terms of ordering of the $Mn^{3+}$ cations in sites as far apart as possible in the ac plane ("Wigner crystal" model) and transverse displacements of the $Mn^{4+}O_6$ octahedra in the c direction. These results are not consistent with a recently proposed model based on transmission electron microscopy (TEM) data in which the $Mn^{3+}O_6$ octahedra are arranged in a stripe pattern ("stripe" model). In particular, the large longitudinal stripe modulation revealed by TEM is not observed, suggesting that the TEM data may not be representative of the bulk sample. Within the framework of the "Wigner crystal" model, the magnetic structure at 1.5 K and the charge-ordered superstructure at 160 K and 1.5 K were refined from the neutron data.

INTRODUCTION

Manganese perovskites (MP), with general formula $A_{1-x}A'_xMnO_3$ ($A$ = La, rare earth, $A'$ = Ca, Sr, Ba..), were first discovered in the nineteen fifties[1], and, soon after, their properties were extensively studied. The early studies on these compounds[1-3] already established the extreme complexity of their structural, magnetic and transport phase diagrams as a function of the formal $Mn^{+4}$ content x (electronic doping), average $A$-site cation size $<r_A>$ and temperature. One of the most interesting aspects of the physics of manganese perovskites is the unusually strong interaction between charge carriers and lattice degrees of freedom, due to the size difference between $Mn^{+4}$ and $Mn^{+3}$, and to the tendency of the latter to lower the octahedral symmetry of its coordination shell, thereby lowering the energy of its $e_g$ electron (Jahn-Teller effect). This strong "electron-phonon" coupling, which can be tuned by varying the electronic doping, electronic bandwidth and disorder, gives rise to a complex phenomenology, in which crystallographic structure, magnetic structure and transport properties are intimately related. The so-called charge-ordered phases represent one of the most intriguing manifestations of these effects in these compounds. Below a certain temperature $T_{CO}$, electronic carriers become localized onto specific sites, which display long-range order throughout the crystal structure (*charge ordering*). Moreover, the filled $Mn^{+3}$-$e_g$ orbitals ($3d_{z^2}$) and the associated lattice distortions (elongated Mn-O bonds) also develop long-range order (*orbital ordering*). Finally, the magnetic exchange interactions between neighboring Mn ions, mediated by oxygen ions, become strongly anisotropic at the local level, since Mn-O-Mn superexchange interactions are *ferromagnetic* through a filled and an empty $3d_{z^2}$ orbital, but *antiferromagnetic* through two empty $3d_{z^2}$ orbitals. This, in turn, gives rise to complex *magnetic ordering* in the structures. Historically, magnetic ordering was the first to be investigated; for example, in the classic 1955 work by Wollan and Koheler [2], the magnetic structures of a series of manganese perovskites with general formula $La_{1-x}Ca_xMnO_3$ ($0 \leq x \leq 1$) were studied using neutron powder diffraction. Immediately afterwards, Goodenough [3] interpreted the large antiferromagnetic superstructure (C-E) found for the composition $La_{0.5}Ca_{0.5}MnO_3$ as evidence of charge ordering, and hypothesized a possible orbital ordering pattern associated with it. It was not until recently, however, that the crystallographic superstructure of $La_{0.5}Ca_{0.5}MnO_3$ arising from charge and orbital ordering was observed by electron diffraction (ED) [4], and subsequently solved by synchrotron x-ray and neutron powder diffraction [5]. More recently, a similar pattern of charge, orbital and magnetic ordering was detected for the composition $Pr_{0.7}Ca_{0.3}MnO_3$, indicating that charge ordering is not limited to the x=0.5 doping level [6]. In $Pr_{0.7}Ca_{0.3}MnO_3$, and, more generally, for all charge-ordered phases with x<0.5, the sizes of the crystallographic and magnetic supercells remain the same as for $La_{0.5}Ca_{0.5}MnO_3$, since a fraction of the $Mn^{+4}$ sites are believed to be randomly occupied by the excess $Mn^{+3}$ ions, with the $3d_{z^2}$ orbital pointing out of the a-c plane where orbital ordering occurs. This mechanism, however, is not available for x>0.5, and charge ordering in this composition range is likely to result in even larger and more complex superstructures.

By employing the same ED and transmission electron microscopy (TEM) techniques used for $La_{0.5}Ca_{0.5}MnO_3$, Chen and coworkers [7,8] have studied a number of other compounds in the $La_{1-x}Ca_xMnO_3$ series ($La_{0.333}Ca_{0.667}MnO_3$, $La_{0.25}Ca_{0.75}MnO_3$), and evidenced the presence of superlattice reflections, indicative of large charge-ordered supercells. In particular, the ED pattern of $La_{0.333}Ca_{0.667}MnO_3$ was found to be consistent with a supercell *tripled* along one of the in-plane crystallographic axes (3a×b×c or a×b×3c). In the real-space images, these superstructures generate a very peculiar, stripe-like contrast. In the original model for the superstructure proposed by Chen *et al.* [7], the $Mn^{+3}$ ions were hypothesized to be located as far apart as possible in order to minimize the Coulomb repulsion energy (Figure 1a). This is reminiscent of the well-know "Wigner crystal" arrangement of charges in low-carrier-density metals [9]. Clearly, the present case is different from the one proposed by Wigner, first of all because the carrier density is quite high, and then because the electron-electron interaction is only one of the factors, the lattice energy being at least as important. For instance, the stacking of equal charges along the b axis, which is common to all models examined herein, clearly arises from the need to minimize the lattice strain associated with orbital ordering. Nevertheless, we will adopt the designation "Wigner crystal" for models where the $e_g$ electrons are located as far apart as possible *within* the a-c plane. The original model for $La_{1-x}Ca_xMnO_3$, based on the "Wigner crystal", was subsequently revised by Mori *et al.* [8]. In the revised model (Figure 1b), $Mn^{+3}$ ions are concentrated in 8.25 Å-wide "bi-stripes", which would have the composition $(La,Ca)_3Mn^{+3}{}_2Mn^{+4}O_9$ and the same orbital-ordered arrangement as $La_{0.5}Ca_{0.5}MnO_3$. In order to yield the overall stoichiometry, these $Mn^{+3}$-rich regions would alternate with $Mn^{+3}$-depleted "background" regions, with composition $(2r-1) \cdot (La,Ca)Mn^{+4}O_3$, r being the $Mn^{+4}/Mn^{+3}$ ratio: $r = \frac{x}{1-x}$. In the case of $La_{0.333}Ca_{0.667}MnO_3$, $(2r-1) = 3$, and the two regions ("bi-stripe" and "background") contain the same number of unit formulas (see Figure 1b). It is noteworthy that, within the framework of this classification scheme, the "Wigner crystal" and "stripe" models are indistinguishable for $La_{0.5}Ca_{0.5}MnO_3$, since they yield the same charge-ordered arrangement. This observation raises an important issue about the x=0.5 composition. In the "stripe" model, for x>0.5 and r integer, the TEM contrast would be associated with a specific topological arrangement in stripes of the $Mn^{+3}/Mn^{+4}$ cations in the lattice. However, for x=0.5, there is a perfect topological symmetry between "bi-stripe" and "background", and an additional (spontaneous) symmetry-breaking mechanism would be required to explain the formation of contrast, which is present at this composition as well. On the contrary, for the "Wigner crystal" model, there is no obvious topological cause to explain the TEM contrast, but neither is there a topological "anomaly" at x=0.5. In this scenario, the TEM contrast at all compositions is likely to be explainable by a single mechanism, whatever this might be.

A careful analysis of the TEM images [8] suggests that "stripe" and "background" regions may have different lattice spacings, indicative of a possible longitudinal modulation (in addition to the transverse modulation found in La$_{0.5}$Ca$_{0.5}$MnO$_3$). This is a further argument in favor of the "stripe" model, since "bi-stripe" and "background" regions are expected to have different lattice spacings, due to the different sizes of the Mn$^{+3}$ and Mn$^{+4}$ ions. Nevertheless, the magnitude of the longitudinal modulation (0.9 Å) is far too large, and is completely incompatible with the average crystallographic structure [5], almost certainly suggesting some kind of aberration in the TEM images.

Distinguishing between the "Wigner crystal" and the "stripe" models is of the utmost importance, not only for manganese perovskites, but also in the context of the broader subject of the physics of transition metal oxides. In fact, stripe-like structural features appear to play a role in determining the physical properties of nickel oxides [10] and copper oxides [11] as well.

In the present paper, we focus attention on the crystallographic and magnetic superstructures of La$_{0.333}$Ca$_{0.667}$MnO$_3$, which we have investigated by means of neutron and x-ray powder diffraction, with the aim of distinguishing between the "Wigner crystal" and the "stripe" models. The antiferromagnetic structure of La$_{0.333}$Ca$_{0.667}$MnO$_3$, as determined from neutron powder diffraction data at 1.5 K, is consistent with a *tripling* of one of the in-plane lattice parameters (a or c, using the orthorhombic Pnma space group setting) and the *doubling* of the other. The overall arrangement of the spins can be interpreted as arising from either a "Wigner crystal" or a "stripe" arrangement of the Mn ions. However, the best fit to the neutron diffraction data is obtained using a non-collinear model, which can be most naturally interpreted in the framework of the "Wigner crystal" model because of the need to reduce spin frustration. Concerning the crystallographic superstructure, the simultaneous analysis of x-ray and neutron diffraction data clearly indicates the presence of a *transverse* modulation, with propagation vector $(2\pi/a)(1/3,0,0)$, the fundamental unit cell being *tripled* in the a-direction. This result contradicts the TEM evidence for a large longitudinal modulation, but, in itself, does not rule out a stripe-like arrangement of the Mn ions, albeit with a different displacement pattern. However, a quantitative comparison between the two models clearly indicates that, within the limits of the sets of constraints imposed, the "Wigner-crystal" model provides a much more satisfactory fit to both neutron and x-ray powder diffraction data. Furthermore, the coordination of the Mn species, as determined from Rietveld refinements based on the "Wigner-crystal" model, is consistent with the sign of the interactions in the magnetic structure. We therefore conclude that the present data provide strong evidence for a Wigner-crystal-like arrangement of the Mn$^{+3}$ and Mn$^{+4}$ ions in La$_{0.333}$Ca$_{0.667}$MnO$_3$.

II. EXPERIMENTAL

The La$_{0.333}$Ca$_{0.667}$MnO$_3$ powder sample was synthesized by a solid-state reaction. Starting materials of La$_2$O$_3$, CaCO$_3$ and MnO$_2$ were mixed in stoichiometric proportions, and heated in air at 1250-1400 °C for three days, with intermediate grindings. The antiferromagnetic transition temperature was determined from magnetization data, as described in Reference [12], and later confirmed by neutron powder diffraction. Neutron powder diffraction data were collected on the high-resolution instrument

D2B at the Institut Laue Langevin in Grenoble, France. Data were collected on *warming* at 1.5 K, 160 K, 220 K, 260 K and 300 K, using a wavelength $\lambda_1 = 1.594$ Å, and at 1.5 K, 220 K and 300 K using a higher resolution configuration and a wavelength $\lambda_2 = 2.400$ Å. The higher-resolution, long-wavelength data were used to index the complex magnetic superstructure, the data at both wavelengths were employed to refine the lattice parameters, and the final structural refinements were carried out based on the 1.594 Å data alone. X-ray powder diffraction data were collected on the X7A beamline at the National Synchrotron Light Source (NSLS) at Brookhaven National Laboratory. A first series of data between 300 K and 50 K was collected on *cooling*, using a flat-plate configuration and a wavelength of 1.1418 Å. These data were used to determine the lattice parameters and the temperature evolution of the superlattice peaks. Subsequently, data were collected at 100 K, 200 K and 260 K using a Debye-Scherrer configuration (0.3 mm capillary) and a wavelength of 0.6941 Å. In both cases, a multiwire proportional linear position-sensitive detector (PSD), operating with a 90%-Xe/10%-$CO_2$ gas mixture at 4 bar was used. The structure refinements were carried out by the Rietveld method using the GSAS[13] and FULLPROF[14] programs.

III. AVERAGE CRYSTAL STRUCTURE

A first indication of the presence of orbital ordering can be gathered from the temperature dependence of the lattice parameters (Figure 2). As in the case of $La_{0.5}Ca_{0.5}MnO_3$, the b-axis rapidly decreases below the charge-ordering temperature, whereas the a and c axes increase. This is consistent with the long Mn-O bonds, associated with filled $Mn^{+3}$-$3d_{z^2}$ orbitals, becoming predominantly oriented in the a-c plane. This is confirmed by refinements of the average crystallographic structure (Table II), based on neutron powder diffraction data at 300 K and 160 K, with "average" Pnma space group symmetry ($a \approx c \approx \sqrt{2}a_p; b \approx 2a_p$, where $a_p$ is the primitive cubic perovskite lattice parameter). Relevant bond lengths and angles are listed in Table II. As in the case of $La_{0.5}Ca_{0.5}MnO_3$, the Mn-O bond lengths at low temperatures are indicative of an apparent *reverse* Jahn-Teller distortion (with four long and two short bonds). This distortion is much smaller at room temperature, and, as we shall see, results from an averaging out of the true orbital ordering pattern in the a-c plane. The analysis of the anisotropic atomic displacement parameters (ADP, or Debye-Waller thermal factors) $u_{ij}$, clearly indicates that, at 160 K the largest atomic displacements with respect to the average atomic positions occur along the c axis, $u_{33}$ being as much as three times larger than the other diagonal elements of the tensor. Below $T_{CO} = 260$ K, additional intensity appears, mainly at high angle, indicative of the presence of charge ordering (see below).

III. MAGNETIC STRUCTURE

Below the Néel temperature ($T_N = 140$ K), a series of strong additional Bragg peaks, occurring mainly at low angles, give a clear indication of the presence of antiferromagnetic (AFM) ordering (Figure 3). In fact, broad diffuse humps near the AFM Bragg peak positions are already present at 160 K (just above $T_N = 140$ K), due to magnetic critical scattering. All the magnetic peaks can be indexed using commensurate propagation vectors, and can be classified as belonging to 3 groups, with the following Miller indices: I) (h,k,l), with k even and h+l=odd. II) (h/3,k,l/2) with k and l odd; III) (0,k,l/2), with k and l odd. A direct comparison with other samples having the same composition shows that the relative intensities of Group II) and Group III) peaks is constant, whereas Group I), comprising only two peaks above background, shows significant variations with respects to the other two. This strongly suggests that Group I) peaks actually belong to a separate magnetic phase. Group I) reflections correspond to a $\sqrt{2}a_p \times \sqrt{2}a_p \times a_p$ magnetic unit cell, where $a_p$ is the primitive cubic perovskite lattice parameter, consistent with the so-called "type C" AFM structure, which is quite ubiquitous at these doping levels. We have chosen to index these reflections on the same unit cell as the crystal structure, but, due to the limited intensity and number of the C-type peaks in the sample, our choice of orientation with respect to the crystallographic axes is rather arbitrary, and needs further verification on a pure C-type sample. The latter two groups, which account for almost all of the observed magnetic superstructure lines, are consistent with a supercell tripled in one of the in-plane directions (a or c) and doubled in the other (3a×b×2c or 2a×b×3c). This evidence strongly suggests a close link between magnetic ordering and charge ordering, since it is perfectly consistent with the superlattice modulation vector $(2\pi/a)(1/3,0,0)$ which has been recently observed by Chen *et al.* by TEM [4].

It is interesting to note that, to the best of our knowledge, the only previous observation of a similar superstructure was reported by Jirak *et al.* in the compound $Pr_{0.3}Ca_{0.7}MnO_3$ (Reference 15). Jirak observed two sets of lines, of which the first is consistent with our Group I) lines (although the interpretation offered by these the authors is not the C-type structure but a more complex non-collinear arrangement). The second set, (h±δ, k, l); Pnma setting, with δ=0.154, was interpreted in terms of an incommensurate "spiral" phase with wave-vector parallel to the [0 0 1] direction, and coincides with Group II). However, the presence of Group III) reflections indicates that their model is not completely correct in this case, since it yields vanishing intensities for this set of peaks. Nevertheless, some of the features of Jirak's model, such as the periodicity and the non-collinearity, are confirmed by our final refinements.

The intensity of the magnetic reflections for Groups II) and III) can be roughly accounted for by a collinear spin arrangement, which is shown in Figure 1, with the spins lying in the a-c plane, and directed predominantly along the a-axis. Also shown in Figure 1 are the proposed charge- and orbitally-ordered arrangements for the "Wigner crystal" and the "stripe" models (Figures 1a, and 1b, respectively). In antiferromagnetic manganese perovskites, a correlation is expected between the sign of the super-exchange interaction between adjacent Mn ions and the occupation of the orbitals that make up the super-exchange path through a single oxygen ion. The interaction is expected to be *ferromagnetic* if a filled $d_{z^2}$ orbital is involved, *antiferromagnetic* otherwise. This correlation was first proposed by Goodenough [3], and is verified completely in the case of $La_{0.5}Ca_{0.5}MnO_3$. However, in the present case, the comparison between charge, orbital and magnetic ordering clearly indicates that in neither case can the aforementioned correlation be satisfied without some degree of magnetic frustration. In other words, some of the bonds (indicated with a wavy line in the figure) have a magnetic interaction of an *opposite sign* with respect to the sign predicted on the basis of the orbital occupation. The number of frustrated bonds is the same for both models (4 in each magnetic unit cell), but their topology is different, with the "stripe" model having a higher degree of frustration associated with the $Mn^{+4}$ ions in the regions between the "stripes".

With the collinear model, the agreement between calculated and observed pattern is quite good. Nevertheless, the presence of some systematic errors suggests that the true spin arrangement is not collinear, and some non-collinear models yield marked improvements in the fits. The refined values of the magnetic moments for the best non-collinear model, based on the *Pm* space group symmetry, are listed in Table I, and the Rietveld refinement profile of the 1.5 K neutron powder diffraction data, based on this magnetic model and an unconstrained refinement of the crystallographic superstructure, is shown in Figure 3. The canted antiferromagnetic structure is shown in Figure 4. Although we cannot claim that this model is unique, it is interesting to examine the physical implications of the canting for the charge-ordered models. Within the framework of the "Wigner crystal model", the interpretation of the canted structure is quite straightforward. The main effect of the canting is to remove part of the frustration on the $Mn^{+4}$-O-$Mn^{+4}$ ferromagnetic (frustrated) interaction, by tilting the angle between the two spins 56 degrees away from collinearity. This is accomplished at the expense of the adjacent $Mn^{+4}$-O-$Mn^{+4}$ *antiferromagnetic* (normal) interaction, where the angle is tilted away from 180 to 124 degrees. On the other hand, no such simple interpretation appears to exist for the "stripe" model. This observation may in itself be an indication that the underlying charge-ordered structure is better represented by the "Wigner crystal" model.

IV. CRYSTALLOGRAPHIC SUPERSTRUCTURE

A series of superlattice reflections are present in all the x-ray and neutron data below $T_{CO}$ = 260 K. The x-ray data are more suitable for indexing these additional peaks, due to the better resolution and to the fact that they are observable also at low angles. The positions of these extra reflections are consistent with a *tripling* of the *a* (not *c*) lattice parameter (propagation vector: $(2\pi/a)(1/3,0,0)$). As an example, the temperature dependence of the integrated intensity of the [2/3,2,3] satellite reflections, as measured by x-ray powder diffraction, is shown in Figure 5. As in the case of $La_{0.5}Ca_{0.5}MnO_3$, the strongest superlattice peaks appear as *satellites* of intense primitive perovskite peaks, with in-phase La and Mn contribution and non-zero l Miller index (e.g., [2/3,2,1], [4/3,2,1], [5/3,0,2], [7/3,0,2], [2/3,2,3], [4/3,2,3], etc.). These observations, and the *c*-axis orientation of the principal axes of the atomic displacement ellipsoids in the average structure (determined from the neutron data), are a sufficiently clear indication that the extra reflections originate from a predominantly *transverse* modulation. Clearly, this is in contrast with the recent TEM results on the same compounds [8], which were interpreted in terms of a large *longitudinal* modulation. Nevertheless, the presence of a transverse modulation does not in itself rule out a stripe-like arrangement of the $Mn^{+3}/Mn^{+4}$ cations. Two plausible models for the displacement patterns associated with the transverse modulation are shown in Figure 6 a) and b), for a "Wigner crystal" and "stripe" arrangement of the Mn ions, respectively. The two models were constructed in such a way as to preserve reasonable bond lengths around each of the two Mn species, and also to comply with the aforementioned selection rules for superlattice peaks. Nevertheless, unlike the magnetic structures, the displacement patterns of the two models are clearly different, and it *must* be possible to distinguish between them from a systematic analysis of the intensities of the superlattice reflections and/or from a full refinement of the superstructures.

One particularly elegant way of distinguishing between the two models is based on Fourier analysis. In the top panels of Figure 6, the positive Fourier components associated with the displacements of the $Mn^{+4}O_6$ octahedra are shown. Even from direct inspection of the two displacement patterns, it is clear that the "Wigner crystal" model is essentially a sine-wave modulation with the periodicity of the unit cell, whereas the "stripe" model contains a significant component (approximately 1/3) having one half the periodicity. We believe that this feature is not due to the particular displacement pattern we assumed, but rather to the intrinsic topology of the two models. For small displacive modulations, it is well known that the intensity of the $n^{th}$-order satellite reflection around a given main Bragg peak is proportional to the *square* of the corresponding Fourier component [16]. Consequently, for the "Wigner crystal" model, we expect to observe only first-order satellites, whereas, for the "stripe" model, second-order satellites with an intensity of roughly 1/10 that of the first-order ones should be present. Since some of the first-order satellites are relatively strong in the x-ray diffraction patterns, some of these higher harmonic peaks should also be observable.

In order to verify this concept, we set out to make a quantitative comparison between the "Wigner crystal" and the "stripe" models. The strategy adopted was to obtain the best possible refinement of each model from the neutron data. Subsequently, we calculated the x-ray diffraction patterns based on these models, and compared them directly with the experimental data, in order to check for systematic absences and for the possible presence of higher-order reflections. This is not so easily done with the present neutron data, since, as already mentioned, only partially overlapped, high-angle superlattice peaks are observable. In practice, however, an additional difficulty arises, due to the very different symmetries of the two models. For the "Wigner crystal" model, a large number of symmetry operators of the original Pnma space group symmetry are preserved, yielding a superstructure with the same nominal space group (Pnma) and only 11 atoms in the asymmetric unit cell. For this model, it is possible to refine all the unconstrained atomic positions, based on neutron diffraction data. Reasonable constraints are needed only for the isotropic thermal parameters (see below). In contrast, the symmetry of the "stripe" model is very low (*Pm*), and only the mirror plane perpendicular to the b-axis is preserved. As a consequence, the "stripe" model comprises 42 independent atoms, and an unconstrained refinement is not possible based on the present data. For the purpose of comparing the two models, we therefore chose to allow only a single atomic shift along the c-axis in both cases, as shown in Fig 6. In both cases, the starting model was the previously refined average crystal structure, and the additional displacements were refined based on neutron data at 160 K, using the program FULLPROF, which is better suited than GSAS to impose complex, symmetry-unrelated sets of constraints. After each refinement cycle with the single modulation parameter, the average structure was once again refined until global convergence was achieved. The average structure determined through this procedure was found to be essentially identical to the one listed in Table I, except for the planar oxygen atoms, which are slightly displaced. The relevant parameters of this comparison are listed in Table IV. It is clearly seen that the constrained "Wigner crystal" model yields a very significant improvement over the average structure model with anisotropic thermal parameters, while having considerably fewer refineable parameters. The "stripe" model also yields improved R-factors, albeit not as good as the "Wigner crystal" model. However, the difference between the two models clearly emerges in the comparison of the measured and calculated x-ray diffraction patterns (Fig 7). In addition to yielding a generally poorer agreement for the first-order superlattice reflections, the "stripe" model generates a few second-order peaks, as expected, but these are not observed. This was verified by obtaining data with longer acquisition times in the regions where second-order reflections were expected. In particular, the second-order doublet (2+2q 0 2)+(1-2q,2,3) (i.e., (8/3 0 2)+(1/3,2,3)), shown in the inset to Fig. 7, is well separated from adjacent reflections and has a significant combined integrated intensity of 84 counts., while the observed intensity is 0(2). Note that the two peaks are only accidentally degenerate, and there is not reason why they should be absent at the same time. Likewise, there is no trace of (2/3,2,3) nor (5/3,2,3). On the contrary, the "Wigner crystal" model generates all the observed satellites and no others, the intensity agreement being quite good. On the basis of this comparison, *we conclude that the charge-ordered arrangement of $Mn^{+3}$ and $Mn^{+4}$ ions in $La_{0.333}Ca_{0.667}MnO_3$ is very likely to be that of a Wigner crystal.*

With this fundamental feature established to the best of our abilities, given the quality of the present data, we proceeded to obtain the best possible refinement of the "Wigner crystal" model. Using FULLPROF, and allowing all the atomic coordinates to vary, consistently with Pnma symmetry, the improvement over the constrained model is rather modest (Table IV), confirming that the displacements are predominantly along the c-axis. A significantly better $R_{wp}$ can be obtained using GSAS, which, we believe, is mainly due to the better modeling of the diffuse background. The structural parameters at 160 K and 1.5 K, as obtained from the latter fits, are listed in Table V. At 1.5 K, the previously described antiferromagnetic structure was also incorporated in the model (see Table III). Relevant bond lengths and angles are listed in Table VI, the refined neutron powder diffraction pattern at 160 K is shown in Fig. 8, and the superstructure is depicted in Fig. 9.

Based on the present model, it is interesting to consider in more detail the coordination of the two Mn species, with particular reference to the previously described magnetic structural model. The $Mn^{+3}$ cations have a classic Jahn-Teller-distorted coordination with the neighboring oxygen atoms, with two long (~2.0 Å) and four short (~1.9 Å) distances. The coordination shell around $Mn^{+4}$ is more unusual, with 5 short bonds (~1.9 Å) and a single long (~2.0 Å) $Mn^{+4}$-O distance. Given the poor resolution of the satellites in the neutron data, the quantitative significance of this result needs to be checked with more accurate (preferably single-crystal) data.. however, it is noteworthy that the single long $Mn^{+4}$-O bond corresponds to the position of the frustrated Mn-O-Mn interaction, and it is not unreasonable that part of this frustration is relieved by a bond length increase. This correspondence is further evidence for the topological correctness of the "Wigner crystal" model.

V. DISCUSSION

The present work, based on the analysis of the crystal and magnetic structures, establishes with a good degree of confidence that the ordering of the $Mn^{+3}$ and $Mn^{+4}$ cations in $La_{0.333}Ca_{0.667}MnO_3$ corresponds to the "Wigner crystal" arrangement, shown in Figure 9. Although the data do not allow us to preclude the transverse "stripe" modulation entirely, we do not believe the evidence supports this model. If confirmed, the "Wigner crystal" model would remove the problem of the topological anomaly at x=0.5, as explained in the introduction, but would totally fail to explain the stripe-like TEM contrast observed by Chen *et al*. Alternatively, for $La_{0.333}Ca_{0.667}MnO_3$, it is highly unlikely but not entirely impossible that a stripe-like arrangement of the Mn ions is associated with another type of transverse displacement pattern, which we might have overlooked, and for which *all* the second-order satellites that we measured are accidentally extinct. We note, however, that this explanation could not apply to $La_{0.5}Ca_{0.5}MnO_3$, for in this case, as already mentioned, the only possible transverse modulation is degenerate with the "Wigner crystal" arrangement. Furthermore, the absence of a longitudinal component has been recently confirmed by very precise synchrotron x-ray diffraction measurements made at the European Synchrotron Radiation Facility, in Grenoble [17]. It is almost certain that dynamic and multiple-scattering effects have significantly enhanced the longitudinal modulation associated with the stripe-like TEM contrast. However, in our opinion, the longitudinal *nature* of the TEM modulation is unlikely to be a complete artifact of the imaging technique. A useful indication in this context is the fact that, with TEM, modulation wave-vectors oriented along *both* in-plane directions (a and c) are observed with roughly equal probability within the *same* orthorhombic grain [7]. This observation cannot be an artifact, and is in contradiction with our data, which clearly show a *single* modulation wave-vector, $(2\pi/a)(1/3,0,0)$. Therefore, a more likely possibility is that a "stripe" type of charge-ordering pattern may occasionally occur in some crystallites, perhaps due to high strains or grain-boundary effects.

However, these regions would not be representative of the bulk, but only account for a minute volume fraction of the entire sample. Nevertheless, we suggest that the structural models described in the present paper and in Reference 5 for $La_{0.333}Ca_{0.667}MnO_3$ and $La_{0.5}Ca_{0.5}MnO_3$ should be used as bases for detailed TEM image simulations, in an attempt to reproduce the observed stripe-like contrast. In fact, if the TEM images were representative of the bulk, the same structural model must be able to account for both the TEM and the diffraction observations.

# Figure Captions

**Figure 1.** Schematic representations of the "Wigner crystal" (a) and "stripe" (b) models for the magnetic, charge and orbital ordered superstructures. $Mn^{+4}$ and $Mn^{+3}$ are shown, respectively, as plain circles and circles with "orbitals". Plusses (+) and minuses (-) indicate the direction of the largest component of the spins (along the orthorhombic a-axis). The actual magnetic structure is non-collinear, with a smaller component along the c-axis (shown in Figure 4, see below). Solid and dotted lines indicate ferromagnetic and antiferromagnetic interactions, respectively. Wavy lines indicate frustrated ferromagnetic interactions, which are expected to be antiferromagnetic based on orbital ordering. The magnetic coupling along the b-axis (perpendicular to the plane of the sheet) is antiferromagnetic for all atoms.

**Figure 2.** Lattice parameters as a function of temperature for $La_{0.333}Ca_{0.667}MnO_3$. Filled and open symbols are used for neutron powder diffraction data (measurement on *warming*) and synchrotron x-ray diffraction data (measurement on *cooling*), respectively. When not shown, error bars are smaller than the symbols.

**Figure 3.** Rietveld plot of $La_{0.333}Ca_{0.667}MnO_3$ neutron powder diffraction data at 1.5 K (l = 1.5943 Å), fitted using the magnetic structure described in the text (see also Table II) and the "Wigner crystal" structural model (see below). The experimental points are indicated as crosses (+). The continuous line is the calculated fit to the pattern. A difference curve is plotted at the bottom. The tick marks indicate Bragg reflections for the following phases (starting from the top): main crystallographic phase (*Pnma*, 3a×b×c); main magnetic structure (3a×b×2c, fitted in the *Pm* space group), associated with Group II and III peaks; "impurity" magnetic structure (Group I peaks). The inset shows an enlargement of the low-angle portion of the profile, with some of the more intense magnetic peaks labeled as described in the text. Nuclear reflections are marked with an "N".

**Figure 4.** Canted antiferromagnetic structure of $La_{0.333}Ca_{0.667}MnO_3$ as refined from neutron powder diffraction data. The values of the magnetic moments are reported in Table III. The arrows indicate the direction of the magnetic moments, while the other symbols are as in Figure 1. Wavy lines are used for the "canted" AFM interaction (see text).

**Figure 5.** Integrated intensity of the [2/3,2,3] superlattice peak as a function of temperature, from synchrotron x-ray diffraction data measured on *cooling*. The appearance of the peak near 160 K marks the onset of the charge-ordering (CO).

**Figure 6.** Displacement patterns for the "Wigner crystal" and "stripe" models (**bottom**) and their Fourier transform (**top**), evidencing the presence of a significant second harmonic component for the "stripe" model. The "Wigner crystal" model is essentially a pure sine modulation.

**Figure 7.** Selected portion of the x-ray powder diffraction pattern for $La_{0.333}Ca_{0.667}MnO_3$ ($\lambda = 0.6941$ Å), enlarged to show some of the more intense intense superlattice reflections (indicated by arrows and dotted lines). The three panels show the same experimental data at 100 K, (shown as plusses, +) and different calculated profiles (solid lines), as calculated from best fits to neutron powder diffraction data at 160 K, using different structural models. a) Average structure (*Pnma*, a×b×c). The superlattice peaks are clearly not accounted for. b) "Stripe" model (*Pm*, 3a×b×c): calculated superlattice peaks are present, but show relatively poor intensity agreement. More importantly, some unobserved second-order reflections (e.g., the doublet (2+2q 0 2)+(1-2q,2,3), marked with an asterisk. *) have significant intensities but are not observed. c) "Wigner crystal" model (*Pnma*, 3a×b×c): all the observed satellite reflections are in good agreement with the calculated, and no second-order peaks are present. The inset show data obtained in the region of the

(8/3 0 2)+(1/3,2,3) doublet with longer acquisition times (the background was subtracted in this case).

**Figure 8.** Rietveld plot of $La_{0.333}Ca_{0.667}MnO_3$ neutron powder diffraction data at 160 K ($\lambda$ = 1.5943 Å), fitted using the "Wigner crystal" structural model. The notation is as in Figure 3, but a single set of tick marks is used, representing the position of the Bragg peaks for the main crystallographic phase (*Pnma*, 3a×b×c). Magnetic diffuse scattering is present at low angle, indicating the incipient antiferromagnetic ordering transition.

**Figure 9.** "Wigner crystal" superstructure of $La_{0.333}Ca_{0.667}MnO_3$. The projection shows the a-c plane. Mn and La atoms are labeled, while oxygen atoms are shown in a darker shade of grey. Some relevant bond lengths in Å are shown.

TABLE I Average $La_{0.333}Ca_{0.667}MnO_3$ structural parameters at 300 K and 160 K, above and below the charge-ordering transition, respectively. The parameters were refined using the program GSAS, based on neutron powder diffraction data at a single wavelength (1.59434 Å). The space group is *Pnma* (No. 62) in both cases, and the structural model has 27 independent internal atomic parameters (coordinates and $U_{ij}$'s). The atomic sites are: La/Ca $4c[x,1/4,z]$; Mn $4a[1/2,0,0]$; $O_a$ $4c[x,1/4,z]$; $O_p$ $8d[x,y,z]$. In order to simplify the notation here and in Table V, the symbols $O_a$ (apical) and $O_p$ (planar) are used instead of O(1) and O(2), as in reference 5. The fractional occupancies were fixed at the nominal stoichiometric values and not refined. Number in parentheses are statistical errors with respect to the last significant digit.

| Parameter | | T=300 K | T=160 K |
|---|---|---|---|
| a (Å) | | 5.3812(1) | 5.3962(1) |
| b (Å) | | 7.5687(1) | 7.4988(1) |
| c (Å) | | 5.3864(1) | 5.4081(1) |
| V (Å$^2$) | | 219.384(9) | 218.842(9) |
| La, Ca | x | 0.0216(3) | 0.0223(3) |
| | z | -0.0037(5) | -0.0031(4) |
| | $U_{11}$ (Å$^2$) | 0.0103(3) | 0.0088(8) |
| | $U_{22}$ (Å$^2$) | 0.0103$^\S$ | 0.0061(5) |
| | $U_{33}$ (Å$^2$) | 0.0103$^\S$ | 0.0160(9) |
| | $U_{13}$ (Å$^2$) | 0$^\S$ | -0.0005(6) |
| Mn | $U_{11}$ (Å$^2$) | 0.0031(3) | 0.007(1) |
| | $U_{22}$ (Å$^2$) | 0.0031$^\S$ | -0.0017(7) |
| | $U_{33}$ (Å$^2$) | 0.0031$^\S$ | 0.008(1) |
| | $U_{12}$ (Å$^2$) | 0$^\S$ | -0.0028(9) |
| | $U_{13}$ (Å$^2$) | 0$^\S$ | -0.0005(3) |
| | $U_{23}$ (Å$^2$) | 0$^\S$ | -0.0010(6) |
| $O_a$ | x | 0.4921(5) | 0.4930(4) |
| | z | 0.0595(3) | 0.0637(3) |
| | $U_{11}$ (Å$^2$) | 0.0170(11) | 0.0059(8) |
| | $U_{22}$ (Å$^2$) | 0.0081(7) | 0.0032(5) |
| | $U_{33}$ (Å$^2$) | 0.0127(8) | 0.0159(8) |
| | $U_{13}$ (Å$^2$) | -0.0031(9) | 0.0029(8) |
| $O_p$ | x | 0.2788(3) | 0.2758(2) |
| | y | 0.0310(1) | 0.0325(1) |
| | z | 0.7231(3) | 0.7251(4) |
| | $U_{11}$ (Å$^2$) | 0.0104(7) | 0.0066(6) |
| | $U_{22}$ (Å$^2$) | 0.0076(4) | 0.0059(4) |
| | $U_{33}$ (Å$^2$) | 0.0121(8) | 0.0221(8) |
| | $U_{12}$ (Å$^2$) | -0.0031(7) | 0.0000(6) |
| | $U_{13}$ (Å$^2$) | -0.0042(4) | 0.0000(5) |
| | $U_{23}$ (Å$^2$) | -0.0034(9) | 0.0035(7) |
| $R_{wp}$(%) | | 5.38 | 5.00 |
| $R(F^2)$(%) | | 3.43 | 3.14 |
| $\chi^2$ (%) | | 3.95 | 5.54 |

$^\S$ For the 300 K data, the atomic displacement parameters for La and Mn were refined isotropically.

TABLE II Selected bond lengths and angles for $La_{0.333}Ca_{0.667}MnO_3$ at 300 K and 160 K, as calculated from the average structure model from Table I. The multiplicities of the bonds associated with each Mn species are listed in the second column.

| Parameter | Multiplicity | T=300 K | T=160 K |
|---|---|---|---|
| $Mn-O_a$ (Å) | ×2 | 1.9196(3) | 1.9065(3) |
| $Mn-O_p$ (Å) | ×2 | 1.936(2) | 1.938(2) |
|  | ×2 | 1.923(2) | 1.932(2) |
| $O_a-Mn-O_a$ (deg) | ×1 | 180 | 180 |
| $O_p-Mn-O_p$ (deg) | ×2 | 180 | 180 |
| $Mn-O_a-Mn$ (deg) | ×2 | 160.6(1) | 159.05(9) |
| $Mn-O_p-Mn$ (deg) | ×4($Mn^{+4}$); ×2($Mn^{+3}$) | 161.13(6) | 161.48(5) |

TABLE III Refined magnetic moment for the antiferromagnetic structure of La$_{0.333}$Ca$_{0.667}$MnO$_3$ at 1.5 K. The fractional atomic coordinates are based on the (3a×b×2c) magnetic supercell. The coupling along the b- and c-axes is antiferromagnetic for all atoms. Following the usual conventions for the choice of origin of the Pm space group, used for the magnetic structure, the coordinates of the Mn ions are shifted by [1/2,1/4,0] with respect to those in Tables I and V.

| Species* | Sites | $\mu_x(\mu_B)$ | $\mu_z(\mu_B)$ |
|---|---|---|---|
| Mn$^{+3}$ | $\left[0,\frac{1}{4},0\right],\left[0,\frac{3}{4},\frac{1}{2}\right]$ | +2.27(8) | +0.6(2) |
|  | $\left[\frac{1}{2},\frac{1}{4},\frac{1}{4}\right],\left[\frac{1}{2},\frac{3}{4},\frac{3}{4}\right]$ | −2.27(8) | +0.6(2) |
|  | $\left[0,\frac{1}{4},\frac{1}{2}\right],\left[0,\frac{3}{4},0\right]$ | −2.27(8) | −0.6(2) |
|  | $\left[\frac{1}{2},\frac{1}{4},\frac{3}{4}\right],\left[\frac{1}{2},\frac{3}{4},\frac{1}{4}\right]$ | +2.27(8) | −0.6(2) |
| Mn$^{+4}$ | $\left[\frac{1}{3},\frac{1}{4},0\right],\left[\frac{1}{3},\frac{3}{4},\frac{1}{2}\right]$ | −1.88(6) | +1.64(6) |
|  | $\left[\frac{2}{3},\frac{1}{4},0\right],\left[\frac{2}{3},\frac{3}{4},\frac{1}{2}\right]$ | +1.88(6) | −1.64(6) |
|  | $\left[\frac{1}{6},\frac{1}{4},\frac{1}{4}\right],\left[\frac{1}{6},\frac{3}{4},\frac{3}{4}\right]$ | −1.88(6) | −1.64(6) |
|  | $\left[\frac{5}{6},\frac{1}{4},\frac{1}{4}\right],\left[\frac{5}{6},\frac{3}{4},\frac{3}{4}\right]$ | +1.88(6) | +1.64(6) |
|  | $\left[\frac{1}{3},\frac{1}{4},\frac{1}{2}\right],\left[\frac{1}{3},\frac{3}{4},0\right]$ | +1.88(6) | −1.64(6) |
|  | $\left[\frac{2}{3},\frac{1}{4},\frac{1}{2}\right],\left[\frac{2}{3},\frac{3}{4},0\right]$ | −1.88(6) | +1.64(6) |
|  | $\left[\frac{1}{6},\frac{1}{4},\frac{3}{4}\right],\left[\frac{1}{6},\frac{3}{4},\frac{1}{4}\right]$ | +1.88(6) | +1.64(6) |
|  | $\left[\frac{5}{6},\frac{1}{4},\frac{3}{4}\right],\left[\frac{5}{6},\frac{3}{4},\frac{1}{4}\right]$ | −1.88(6) | −1.64(6) |

* The species assignment and the final constraints on the magnetic moments are based on the "Wigner crystal" model.

TABLE IV Comparison between average structure, "Wigner crystal" and "stripe" models for $La_{0.333}Ca_{0.667}MnO_3$ at 160 K. In the present comparison, all the refinements were performed using the program FULLPROF, which is better suited to handle the complex constraints of the "stripe" model. In the average structure models, isotropic and anisotropic atomic displacement parameters (ADP) were used, respectively. In the two constrained models, in addition to the isotropic ADP, only a single displacement parameter along the c-axis was refined, as in Figure 6. The other atomic coordinates were fixed at the values for the average structure.

|  | Average Structure (Isotropic ADP) | Average Structure (Anisotropic ADP) | Constrained "Wigner crystal" model | Constrained "stripe" model | Unconstrained "Wigner crystal" model (cfr. Table IV) |
|---|---|---|---|---|---|
| No. of fitted parameters | 29 | 45 | 23 | 23 | 46 |
| No. of atomic parameters | 11 | 27 | 5 | 5 | 28 |
| $R_{wp}$(%) | 7.69 | 6.85 | 6.08 | 6.04 | 5.58 |
| $R(F^2)$(%) | 7.42 | 5.72 | 4.89 | 5.00 | 4.29 |
| $\chi^2$ (%) | 13.2 | 10.7 | 8.20 | 8.56 | 7.01 |

TABLE V La$_{0.333}$Ca$_{0.667}$MnO$_3$ structural parameters at 160 K and 1.5 K, as refined using the "Wigner crystal" model. The program GSAS was used to refined the parameters, based on neutron powder diffraction data at a single wavelength (1.59434 Å). The space group is *Pnma* (No. 62), and the model has 28 independent internal atomic parameters (coordinates and U$_{iso}$'s). The atomic sites are: La/Ca 4*c*[x,1/4,z]; Mn+$^4$ 8*d*[x,y,z]; Mn$^{+3}$ 4*a*[1/2,0,0]; O$_a$ 4*c*[x,1/4,z]; O$_p$ 8*d*[x,y,z]. The fractional occupancies were fixed at the nominal stoichiometric values and not refined. Number in parentheses are statistical errors with respect to the last significant digit. The refinement of the 1.5 K data also includes the antiferromagnetic structure, (see Table III).

| Parameter | | T=160 K | T=1.5 K |
|---|---|---|---|
| a (Å) | | 16.1869(5) | 16.1847(4) |
| b (Å) | | 7.4987(2) | 7.4885(1) |
| c (Å) | | 5.4080(2) | 5.4075(1) |
| V (Å$^2$) | | 656.42(4) | 655.38(2) |
| La/Ca1 | x | 0.0081(4) | 0.0100(4) |
| | z | -0.001(1) | -0.005(1) |
| | U$_{iso}$ (Å$^2$)$^†$ | 0.0074(3) | 0.0050(3) |
| La/Ca2 | x | 0.3403(4) | 0.3405(4) |
| | z | 0.0106(8) | 0.0153(9) |
| | U$_{iso}$ (Å$^2$)$^†$ | 0.0074(3) | 0.0050(3) |
| La/Ca3 | x | 0.6738(4) | 0.6724(4) |
| | z | -0.0216(10) | -0.024(1) |
| | U$_{iso}$ (Å$^2$)$^†$ | 0.0074(3) | 0.0050(3) |
| Mn$^{+4}$ | x | 0.1659(6) | 0.1654(7) |
| | y | 0.002(1) | 0.0000(1) |
| | z | 0.008(1) | 0.012(1) |
| | U$_{iso}$ (Å$^2$)$^†$ | 0.0034(3) | 0.0026(4) |
| Mn$^{+3}$ | U$_{iso}$ (Å$^2$)$^†$ | 0.0034(3) | 0.0026(4) |
| O$_a$1 | x | 0.1639(4) | 0.1637(5) |
| | z | 0.0781(9) | 0.079(1) |
| | U$_{iso}$ (Å$^2$)$^†$ | 0.0058(3) | 0.0051(3) |
| O$_a$2 | x | 0.5015(4) | 0.5012(4) |
| | z | 0.0597(9) | 0.062(1) |
| | U$_{iso}$ (Å$^2$)$^†$ | 0.0058(3) | 0.0051(3) |
| O$_a$3 | x | 0.8260(4) | 0.8268(5) |
| | z | 0.0492(9) | 0.047(1) |
| | U$_{iso}$ (Å$^2$)$^†$ | 0.0058(3) | 0.0051(3) |
| O$_p$1 | x | 0.0919(3) | 0.0938(4) |
| | y | 0.0310(7) | 0.0293(7) |
| | z | 0.7425(8) | 0.7441(8) |
| | U$_{iso}$ (Å$^2$)$^†$ | 0.0073(2) | 0.0060(3) |
| O$_p$2 | x | 0.0752(3) | 0.0759(4) |
| | y | -0.0349(7) | -0.0357(7) |
| | z | 0.2329(8) | 0.2327(9) |
| | U$_{iso}$ (Å$^2$)$^†$ | 0.0073(2) | 0.0060(3) |
| O$_p$3 | x | 0.2398(3) | 0.2412(3) |
| | y | -0.0312(6) | -0.0326(7) |
| | z | 0.3027(5) | 0.3033(6) |
| | U$_{iso}$ (Å$^2$)$^†$ | 0.0073(2) | 0.0060(3) |
| R$_{wp}$(%) | | 4.33 | 5.41 |
| R(F$^2$)(%) | | 2.71 | 3.12 |
| $\chi^2$ (%) | | 4.14 | 3.22 |

$^†$ The isotropic atomic displacement parameters U$_{iso}$ were constrained to be equal for atoms that are equivalent in the average structure (La/Ca, Mn, O$_a$, O$_p$).

TABLE VI Selected bond lengths and angles for $La_{0.333}Ca_{0.667}MnO_3$ at 160 K and 1.5 K, as calculated from the using the "Wigner crystal" model from Table V. The multiplicities of the bonds associated with each Mn species are listed in the second column.

| Parameter | Multiplicity | T=160 K | T=1.5 K |
|---|---|---|---|
| $Mn^{+3}$-$O_a2$ (Å) | ×2 | 1.9024(8) | 1.902(1) |
| $Mn^{+3}$-$O_p1$ (Å) | ×2 | 1.997(5) | 2.024(5) |
| $Mn^{+3}$-$O_p2$ (Å) | ×2 | 1.907(5) | 1.916(6) |
| $Mn^{+4}$-$O_a1$ (Å) | ×1 | 1.900(8) | 1.904(8) |
| $Mn^{+4}$-$O_a3$ (Å) | ×1 | 1.917(8) | 1.907(8) |
| $Mn^{+4}$-$O_p1$ (Å) | ×1 | 1.883(9) | 1.87(1) |
| $Mn^{+4}$-$O_p2$ (Å) | ×1 | 1.93(1) | 1.89(1) |
| $Mn^{+4}$-$O_p3$ (Å) | ×1 | 1.900(9) | 1.90(1) |
| $Mn^{+4}$-$O_p3$ (Å) | ×1 | 2.007(7) | 2.010(9) |
| $O_a1$-$Mn^{+4}$-$O_a3$ (deg) | ×1 | 176.4(6) | 176.8(8) |
| $O_p1$-$Mn^{+4}$-$O_p3$ (deg) | ×1 | 177.1(5) | 179.291(8) |
| $O_p2$-$Mn^{+4}$-$O_p3$ (deg) | ×1 | 176.2(5) | 177.2(6) |
| $O_a2$-$Mn^{+3}$-$O_a2$ (deg) | ×1 | 180 | 180 |
| $O_p1$-$Mn^{+3}$-$O_p1$ (deg) | ×1 | 180 | 180 |
| $O_p2$-$Mn^{+3}$-$O_p2$ (deg) | ×1 | 180 | 180 |
| $Mn^{+4}$-$O_a1$-$Mn^{+4}$ (deg) | ×1 | 157.0(5) | 158.1(5) |
| $Mn^{+4}$-$O_a3$-$Mn^{+4}$ (deg) | ×1 | 159.7(5) | 158.9(5) |
| $Mn^{+3}$-$O_a2$-$Mn^{+3}$ (deg) | ×2 | 160.4(3) | 159.8(3) |
| $Mn^{+4}$-$O_p1$-$Mn^{+3}$ (deg) | ×1($Mn^{+4}$); ×2($Mn^{+3}$) | 164.0(4) | 163.5(4) |
| $Mn^{+4}$-$O_p2$-$Mn^{+3}$ (deg) | ×1($Mn^{+4}$); ×2($Mn^{+3}$) | 160.9(4) | 160.8(4) |
| $Mn^{+4}$-$O_p3$-$Mn^{+4}$ (deg) | ×2 | 158.2(3) | 159.1(3) |